\documentstyle[amsfonts,preprint,aps]{revtex}
\tightenlines

\begin{document}
\title{Angular momentum effects in weak gravitational fields }
\author{A. Tartaglia$^{\thanks{%
Permanent address: Dip. Fisica, Politecnico, Torino, Italy}}$}
\address{Gravity Probe B, Hansen Experimental Physics Labs, Stanford University,\\
Stanford (CA) and INFN, Torino, Italy\\
e-mail: tartaglia@polito.it}
\date{\today }
\maketitle
\pacs{04.25.Nx, 95.30.Sf }

\begin{abstract}
It is shown that, contrary to what is normally expected, it is possible to
have angular momentum effects on the geometry of space time at the
laboratory scale, much bigger than the purely Newtonian effects. This is due
to the fact that the ratio between the angular momentum of a body and its
mass, expressed as a length, is easily greater than the mass itself, again
expressed as a length.
\end{abstract}

\section{Introduction}

The influence of the angular momentum on the gravitational interaction has a
long story, dating back to Newton's rotating bucket. It went through the
attempts to implement Mach's principle and was considered during the XIX
century in the attempt to establish a full correspondence between the
gravitational field and the Maxwell theory of electromagnetism (see for
instance Heaviside \cite{heaviside}). In fact it was only Einstein's general
relativity theory that succeeded in accounting fully for the angular
momentum effects. Since the very beginning a couple of papers by Lense and
Thirring \cite{lense} established the formalism and showed what could be
expected, but at that time was practically unobservable, in the surroundings
of the Earth. Since Lense and Thirring pioneering work an impressive mass of
papers has been produced studying the general relativistic effects of
rotation. The investigation has touched both extremely relativistic
situations such as the neighborhood of black holes or in general hugely
massive objects, and the weak field limit which is expected to be acceptable
in almost any situation within the solar system.

In weak field conditions almost all works have been studying the so call
gravitomagnetic effects, which correspond to the (weak field) decomposition
of the field in a gravito-electric part (Newtonian approximation) and a
gravito-magnetic part, depending on the angular momentum of the central body
much as in the case of a magnetic dipole originated by a closed electric
current loop \cite{gravimag}. The most famous gravitomagnetic effect is
precisely the Lense-Thirring drag inducing a precession on a freely falling
gyroscope just as it would happen for a magnetic dipole moving in the field
of a bigger one.

The actual detection of the Lense-Thirring effect is entrusted both to the
observation of astronomical phenomena, such as the behavior of massive
binary systems or, closer to us, the orbital motion of Earth satellites, and
to direct measurement. Indeed Ciufolini and collaborators \cite{ciufolini}
found out the effect studying the precession of the orbit of LAGEOS\
satellites. A direct experiment performed considering the precession of four
gyroscopes in a polar orbit around the Earth is about to fly in the Gravity
Probe B (GPB)\ program (a collaboration between the Stanford University and
the NASA) \cite{gpb}.

Another possibility is to search for gravitomagnetic clock effects, which
should show up as asymmetries in the time of flight of light moving in
opposite directions around the Earth \cite{mashtart}.

All these effects, as said, belong to the category of gravitomagnetic
effects. These in turn stem out of the same off diagonal term of the metric
in the vicinity of a rotating weakly gravitating body and are describable as
been due to a vector potential, which is in fact proportional to the angular
momentum $\overrightarrow{J}$ of the body. Nobody considered up to now
effects possibly due to higher order terms (in the sense of being dependent
on powers of $J$ higher than the first). The intuitive reason is that since
the Lense-Thirring effect in the terrestrial environment is extremely small,
any second or higher order effect should be negligibly smaller.

This letter will precisely show that this is not the case and that there are
situations where the Newtonian effect is absolutely negligible whereas
second order effects of the rotation are not. This possibility was
inadvertently foreshadowed in \cite{tartaglia} but will now be proved and
explained in the next section, just considering actual numerical values. It
will be clear that there are corrections to the diagonal terms of the metric
tensor that are indeed proportional to the square of the angular speed of
the source of the field and produce effects whose size makes them fit for a
laboratory verification.

\section{Comparison of mass and angular momentum contributions to the metric
tensor}

We shall assume a weak gravitational field context. By weak gravitational
field we mean a situation in which the gravitational potential expressed as
a dimensionless quantity is much less than $1$. Assuming for simplicity a
spherical symmetry the dimensionless Newtonian potential at a distance $r$
from the center of the source is
\[
\varepsilon =\frac{U}{c^{2}}=G\frac{M}{c^{2}r}=\frac{\mu }{r}
\]
The symbols have the usual meaning, $U$ is the Newtonian potential, $\mu $
is the mass of the body measured in meters.

The weak field condition is then
\[
\varepsilon <<1
\]
Considering a rotating isolated body there are in fact two conserved
quantities to be considered in order to describe its effects versus the
surrounding space time: one is of course the total mass $M$, the other is
the total angular momentum ${\bf J}$ (actually here we always use its
projection on the rotation axis). In order to compare the contribution of
both to the space time metric one should construct out of them equally
dimensioned parameters and dimensionless quantities to be confronted with
unity. In the case of the mass term we already have $\mu $ and $\varepsilon $%
. In the case of the angular momentum the length that can be obtained from
it, expressing in a sense the pure rotation, is
\[
a=\frac{J}{Mc}
\]
The corresponding dimensionless quantity is
\[
\alpha =\frac{a}{r}
\]
The parameter $a$ is precisely the same as the one entering the Kerr metric.

Kerr's is indeed the most famous axially symmetric stationary metric. It has
been obtained as an exact solution of the Einstein equations and describes
the space time around a ring singularity \cite{kerr}. Studying Kerr space
times people have virtually considered all possibilities. When it is $\mu >a$
two limiting ordinary surfaces exist, one of which is properly a horizon
\cite{straumann}. When on the contrary $\mu <a$ no horizon exists and one is
confronted with a naked singularity. Outside of the black holes physics it
is usually thought that Kerr metric is of no particular use, also because no
internal solutions to the Einstein equations have by now been found matching
the vacuum Kerr solution. Similarly it is not expected that in ordinary
situations the condition $\mu <a$ can have any meaning.

However it is trivial to show that the $\mu <a$ condition is not at all rare
or unachievable. Let us consider first the case of the Earth: its mass
(expressed in meters) is $\mu _{\oplus }=4.4\times 10^{-3}$ m. To calculate $%
a$ it is convenient to assume the simplifying hypothesis that the body is
spherical, homogeneous and rigidly rotating; it is thus simple from the very
definition of $a$ to obtain
\begin{equation}
a=\frac{2}{5}\frac{R^{2}}{c}\Omega  \label{a-value}
\end{equation}

Here $R$ is the radius of the sphere and $\Omega $ is its angular velocity.
Introducing the numbers for the Earth ($R_{\oplus }=6.3\times 10^{6}$ m, $%
\Omega _{\oplus }=7.3\times 10^{-5}$ s$^{-1}$) we see that
\[
a_{\oplus }=3.86\text{ m}
\]
For the Earth $a$ is almost three orders of magnitude bigger than $\mu $.

If we repeat the exercise for the Sun we find that $\mu _{\odot }=1.48\times
10^{3}$ m and $a_{\odot }\simeq 2\times 10^{3}$ m: $\mu $ and $a$ have the
same order of magnitude.

Why then the Lense-Thirring effect is so small as compared to the Newtonian
and in general gravito-electric effects of the field? The reason is simple:
the typical form of the gravito-magnetic dipole potential, responsible for
the Lense-Thirring effect, is:
\[
{\frak V}=\frac{\overrightarrow{J}\cdot \widehat{r}}{r^{2}}
\]
This quantity is proportional to $\mu a/r^{2}$ i.e. to $\varepsilon \alpha $%
. Considering the surface of the Earth ($r=R_{\oplus }$) it is $\varepsilon
\sim 10^{-8}$ and $\alpha \sim 10^{-6}$. The consequence is that of course
the product of $\alpha $ times $\varepsilon $ is six orders of magnitude
smaller than $\varepsilon $ itself.

\section{Relevance of second or higher order terms}

In the previous section we compared $\varepsilon $ and $\alpha $, however in
the metric of space time surrounding a rotating body higher order terms
should in principle be considered too. The general form of the line element
in the axially symmetric stationary case may be written as
\[
ds^{2}=g_{00}d\tau ^{2}+g_{rr}dt^{2}+g_{\theta \theta }d\theta
^{2}+2g_{0\phi }d\tau d\phi +g_{\phi \phi }d\phi ^{2}
\]
with all $g_{\mu \nu }$ not depending on $\tau $ and $\phi $.

In our weak field conditions it is reasonable to develop the elements of the
metric tensor in powers of the $\varepsilon $ and $\alpha $ quantities (in
practice: in inverse powers of $r$) starting from the flat space time
Lorentz metric. Furthermore since the line element must be even with respect
to time reversal and $a$ (which contains the angular momentum) is odd, we
conclude that the diagonal elements of the metric can contain any power of $%
\varepsilon $, but even powers of $\alpha $ only. As for the off diagonal
term, which multiplies the time differential, it must be odd versus time
reversal; this means that $g_{0\phi }$ can contain no isolated power of $%
\varepsilon $ and only odd powers of $\alpha $. A linear dependence on $%
\alpha $ alone can be eliminated by a simple coordinate transformation, so
the leading term of the development must be proportional to $\alpha
\varepsilon $.

All this is to say that $\alpha ^{2}$ contributions must be considered, then
it is useful to compare their relative size with the one of the $\varepsilon
$ terms.

Returning again to the simplifying description of the rotating homogeneous
sphere one has
\begin{equation}
\varepsilon =G\frac{M}{rc^{2}}=\frac{4}{3}\pi \rho \frac{G}{c^{2}}\frac{R^{3}%
}{r}=\kappa \rho \frac{R^{3}}{r}  \label{eps-gen}
\end{equation}
where $\rho $ is now the (average) density of the sphere. The numerical
factor contained in the parameter $\kappa $ accounts for the actual shape of
the rotating body (in general $R$ would be the radius at the 'equator' of
the body); what matters here is only the order of magnitude of $\kappa $,
which, for a solid body, is
\[
\kappa \sim 10^{-27}\text{ m}\times \text{kg}^{-1}
\]
If a thin walled hollow object was assumed, than $\kappa $ would be rescaled
by the factor $l/R$, where $l$ is the thickness of the shell.

As for $\alpha $ its expression may be recast as
\begin{equation}
\alpha =\xi \frac{R^{2}}{r}\Omega  \label{a-two}
\end{equation}
where again the actual numerical value of $\xi $ depends on the shape of the
object, but the important feature is the order of magnitude that is
\[
\xi \sim 10^{-10}\text{ s/m}
\]
A further remark regarding $a$ is that $v=\Omega R$ represents the maximum
peripheral speed of the rotating body. Consideration of this parameter sets
an upper limit to the allowed values of $\Omega $. Actually the object
should not explode under the action of centrifugal forces. Just to fix ideas
and orders of magnitude we can refer to the fact that the best available
materials \cite{materials} can resist peripheral speeds as high as $v_{\max
}\sim 1000$ m/s. In any case it is convenient to explicitly introduce $v$ in
the $\alpha $ formula:
\begin{equation}
\alpha =\xi v\frac{R}{r}  \label{alfa}
\end{equation}

Returning for a moment to the comparison of $\varepsilon $ with $\alpha $,
looking for the fulfillment of the condition
\begin{equation}
\varepsilon <\alpha  \label{cond-0}
\end{equation}
we see that the mass term is smaller when the radius of the body is
\begin{equation}
R<\sqrt{\frac{\xi v}{\kappa \rho }}  \label{R-cond}
\end{equation}
In 'ordinary' situations, where by 'ordinary' I consider average densities
comparable to the density of water ($\rho \sim 10^{3}$ kg/m$^{3}$) and
peripheral velocities not greater than $v_{\max }$, it is
\begin{equation}
R<\sim 10^{8}\text{ m}  \label{linear}
\end{equation}
Of course slow rotation reduces the value of the upper limit, whereas the
inclusion of self-gravitational effects increases it. In practice however we
see that, letting stars apart, (\ref{cond-0}) is most often satisfied.

Let us now pass to the comparison of $\varepsilon $ with $\alpha ^{2}$.
Considering (\ref{eps-gen}) and (\ref{alfa}) we see that, outside the
rotating body, the region where the latter is greater than the former is
defined by the condition
\begin{equation}
R<r<\frac{\xi ^{2}v^{2}}{\kappa \rho R}  \label{cond-sq}
\end{equation}
(\ref{cond-sq}) has solutions when
\begin{equation}
R<\frac{\xi v}{\sqrt{\kappa \rho }}\sim 10^{5}\text{ m}  \label{square}
\end{equation}
Again the numerical estimate refers to the 'ordinary' situation defined
above.

The next step will be the comparison of $\varepsilon $ with the third power
of $\alpha $. Repeating the scheme outlined before one sees that the region
where $\varepsilon <\alpha ^{3}$ corresponds to
\begin{equation}
R<r<\sqrt{\frac{\xi ^{3}v^{3}}{\kappa \rho }}\sim 10^{2}\text{ m}
\label{cond-three}
\end{equation}
We are visibly approaching laboratory scales.

The last meaningful comparison is between $\varepsilon $ and $\alpha ^{4}$.
Now the region where the mass term continues to be smaller is given by
\begin{equation}
R<r<\sqrt[3]{\frac{\xi ^{4}v^{4}R}{\kappa \rho }}\sim \sqrt[3]{R}
\label{cond-four}
\end{equation}
that can be satisfied only if $R<5$ cm.

Once the scale of the laboratory has been reached it turns out that a thin
shell is more convenient that a solid body. Supposing that the thickness to
radius ratio be $\sim 10^{-3}$ the $\kappa $ parameter has now the value $%
\sim 10^{-30}$ m$\times $kg$^{-1}$. Viceversa considering for instance a
rotating spherical hull the calculation of $a$ produces a factor of $2/3$
instead of the $2/5$ of formula (\ref{a-value}), so the order of magnitude
remains the same. Introducing these changes, the various limits in (\ref
{linear}), (\ref{square}), (\ref{cond-three}) and (\ref{cond-four}) change
too. In particular the upper limits for the third and fourth order
conditions to hold become respectively $\sim 10^{3}$ m and $\sim 0.5$ m.

Summing up the results of this section we conclude that in laboratory
conditions it can easily be

\begin{equation}
\alpha ^{4}<\varepsilon <<\alpha ^{3}<<\alpha ^{2}<<\alpha <<1
\label{gen-cond}
\end{equation}

\section{Conclusion}

Considering (\ref{gen-cond}), the line element in the space time sorrounding
an appropriately rotating body in the laboratory will be written
\begin{equation}
ds^{2}=\left( 1+B_{0}\alpha ^{2}\right) d\tau ^{2}-\left( 1+B_{r}\alpha
^{2}\right) dr^{2}-\left( 1+B_{\theta }\alpha ^{2}\right) r^{2}d\theta
^{2}-\left( 1+B_{\phi }\alpha ^{2}\right) r^{2}\sin ^{2}\theta d\phi ^{2}
\label{new-line}
\end{equation}
The $B$'s are of order $\sim 1$ and depend at most on $\theta $. The off
diagonal term has been dropped because it is of order $\varepsilon \alpha .$

We conclude that, if any effect of spinning bodies on the space time can be
found in a laboratory, it depends on pure rotation.

Of course one could wonder at this point whether in any case the corrections
in (\ref{new-line}) are big enough to produce detectable effects. The answer
is yes. Considering for instance a spherical hull, 1 m in radius, with a 1
mm thick wall, rotating with a peripheral speed of 1000 m/s, its $\alpha $
value on the surface would be
\begin{equation}
\alpha \simeq 10^{-6}  \label{simple}
\end{equation}
Gravitational effects within the solar system, using dimensionless
quantities, are expressed by an $\varepsilon \sim 10^{-8}$. Of course now we
must look at the square of (\ref{simple}) which is four orders of magnitude
smaller than the given $\varepsilon $. However if we consider now the
Lense-Thirring effect on the surface of the Earth we see, recalling the
numbers cited at the end of the introductory section, that, in dimensionless
units, it is of the order of $10^{-14}$. Rotation effects in the laboratory
could be easier to be measured than the Lense-Thirring effect of the whole
Earth.

This opens the way to the possibility of extremely interesting laboratory
scale experiments.

\section{Acknowledgement}

The author wishes to thank the GPB group of the Stanford University for kind
hospitality and for financial support within the GPB program during the
elaboration of the present paper and is particularly grateful to Francis
Everitt, Ron Adler, Alex Silbergleit and Bob Wagoner for many stimulating
discussions.

\end{document}